\documentclass[10pt,a4paper]{article}
\title{\bf Inverse Scattering Method II.\\ {\it Methodological part with an example: Breather solution of the Sine-Gordon Equation.}}
\author{Jana Tothova\\ {\it Lab,} Stierova 23, SK - 040 23  Kosice, Slovak Republic
	\and
	Matej Hud\'{a}k \\ {\it Lab,} Stierova 23, SK - 040 23  Kosice, Slovak Republic \footnote{Corresponding author}\\
	{hudakm@mail.pvt.sk}}
\date{May 13, 2018}

\begin{document}
	\maketitle{}

	\vspace{1in}
	\begin{abstract}
Recently it was discussed the Inverse Scattering Method, Part I. (paper I.) . It was a methodological Part with an example - soliton (kink) solution of the Sine-Gordon Equation. The aim of the paper I. was to introduce the Inverse Scattering Method for later studies of some problems in nonlinear dynamics. As a methodological example we described how to solve the Sine-Gordon Equation using the Inverse Scattering Method to obtain a soliton. This soliton solution is well known. In this paper we discuss another methodological example: the solution of the Sine-Gordon Equation using the Inverse Scattering Method to obtain description of a breather. While this breather solution is also well known, we will discuss later breather solutions in some physical systems. Thus we will have in the Part I. and in this Part II. seeds for solving problems with the Inverse Scattering Method, and  in case of the Sine-Gordon Equation  solutions describing the soliton and the breather.	
	\end{abstract}

	\tableofcontents{}

\newpage
\section{Introduction.}
Recently we discussed \cite{MJO1} the Inverse Scattering Method, part I. . This was a methodological part with an example - soliton solution of the Sine-Gordon Equation. Let us denote here a reference to this paper as I. . 
The aim of the paper I. was to introduce the Inverse Scattering Method for later studies of some problems in nonlinear dynamics. As a methodological example we described the kink solution of the Sine-Gordon Equation using the Inverse Scattering Method. This soliton solution is well known.
In this paper we discuss as another methodological example the breather solution of the Sine-Gordon Equation using the Inverse Scattering Method. This breather solution is also well known. In the next part we discuss the Sine-Gordon Equation and its breather solution. In the last section there is a short discussion.

\section{Sine-Gordon Equation and its Breather Solution.}
Let us choose in Part I. (solution of the Sine-Gordon Equation) as discrete eigenvalues the following pair (denotation of variables is that from the Part I.): $\zeta = i. \eta - \frac{1}{2} \omega $ and $-\zeta^{*} = i. \eta + \frac{1}{2} \omega,  $ here $\omega >0$. We see that $\mid \zeta \mid^{2} = \eta^{2} + \frac{\omega^{2}}{4}$. From this we obtain $\omega^{2} = 4. (\mid \zeta \mid^{2}- \eta^{2})$. Because in our linear problem we have reflexion-less potential the coefficient $b$ will be chosen to be zero $b=0$.   
   
From the system of equations we obtain:
\begin{itemize}
	\item 
	from $b = 0$ we obtain $c(x, \xi)=0$
	\item from the equation $(63)$ in the paper I. we obtain $\Phi_{1}(\xi, x)=0$ 
	\item from the equation $(63)$ in the paper I.  we obtain $\Phi_{2}(\xi, x)=0$ 
	\item from equations $(64)$ and $(65)$ in the paper I.  we obtain:
	\begin{equation}\label{1}
	\psi_{1}(x, \zeta_{1}). \exp(-i.\zeta_{1}.x) + \frac{\exp(-i.\zeta_{1}^{*}.x)}{\zeta_{1} - \zeta_{1}^{*}}.c_{1}^{*}. \psi_{2}^{*}(x, \zeta_{1}) + \frac{\exp(-i.\zeta_{2}^{*}.x)}{\zeta_{1} - \zeta_{2}^{*}}.c_{2}^{*}. \psi_{2}^{*}(x, \zeta_{2})=0
	\end{equation} 
	and
\begin{equation}\label{2}
\psi_{1}(x, \zeta_{2}). \exp(-i.\zeta_{2}.x) +
\frac{\exp(-i.\zeta_{1}^{*}.x)}{\zeta_{2} - \zeta_{1}^{*}}.c_{1}^{*}. \psi_{2}^{*}(x, \zeta_{1}) + \frac{\exp(-i.\zeta_{2}^{*}.x)}{\zeta_{2} - \zeta_{2}^{*}}.c_{2}^{*}. \psi_{2}^{*}(x, \zeta_{2})=0
\end{equation}
and
\begin{equation}\label{3}
\psi_{2}^{*}(x, \zeta_{1}). \exp(i.\zeta_{1}^{*}.x) 
- \frac{\exp(i.\zeta_{1}.x)}{\zeta_{1}^{*} - \zeta_{1}}.c_{1}. \psi_{1}^{*}(x, \zeta_{1}) - \frac{\exp(i.\zeta_{2}.x)}{\zeta_{1}^{*}-\zeta_{2}}.c_{2}. \psi_{1}(x, \zeta_{2})=1
\end{equation}
and
\begin{equation}\label{4}
\psi_{2}^{*}(x, \zeta_{2}). \exp(i.\zeta_{2}^{*}.x) 
- \frac{\exp(i.\zeta_{1}.x)}{\zeta_{2}^{*} - \zeta_{1}}.c_{1}. \psi_{1}(x, \zeta_{1}) - \frac{\exp(i.\zeta_{2}.x)}{\zeta_{2}^{*}-\zeta_{2}}.c_{2}. \psi_{1}(x, \zeta_{2})=1.
\end{equation}
\end{itemize}
In equations (\ref{1}) - (\ref{4}) there are constants $c_{1}$ and $c_{2}$ which are unspecified in the time $t = 0$, however their time development is known:
\begin{equation}\label{5}
c_{1} = \exp(- \frac{i.t}{2.\zeta_{1}}) . c_{1}(t=0)
\end{equation}
\[ c_{2} = \exp(- \frac{i.t}{2.\zeta_{2}}) . c_{2}(t=0) .\]
Discrete eigenvalues $\zeta_{1} $ and $\zeta_{2} $ are:
\begin{equation}\label{6}
\zeta_{1} \equiv \zeta = i. \eta - \frac{1}{2} \omega
\end{equation}
\[ \zeta_{2} \equiv -\zeta^{*} = i. \eta + \frac{1}{2} \omega .\]
From equations (\ref{1}) - (\ref{4}) it is necessary to find unknown variables (defined here as $x_{1}$, $x_{2}$, $x_{3}$ and $x_{4}$):
\begin{equation}\label{7}
x_{1} \equiv \psi_{1}(x, \zeta_{1})
\end{equation} 
and
\[ x_{2} \equiv \psi_{1}(x, \zeta_{2})\]
and
\[ x_{3} \equiv \psi_{2}^{*}(x, \zeta_{1}) \]
and
\[ x_{4} \equiv \psi_{2}^{*}(x, \zeta_{1}) .\]
The variable $x$ in the argument of  Jost functions is fixed now, and also $t$ variable! Equations (\ref{1}) - (\ref{4}) are an algebraic system of four equations with four unknown $x_{1}$, $x_{2}$, $x_{3}$ and $x_{4}$ with nonzero right hand side.
Finding unknown $x_{1}$, $x_{2}$, $x_{3}$ and $x_{4}$ we find the potential $q(x, t)$, because according to (\ref{1}) and (\ref{7}) we have:
\begin{equation}\label{8}
q(x, t) = - 2.i.(c_{1}^{*} . \exp(-i .\zeta_{1}^{*}.x).x_{3} + c_{2}^{*} . \exp(-i. \zeta_{2}^{*}.x).x_{4}).
\end{equation}

The system of equations (\ref{1}) - (\ref{4}) can be written in the form:
\begin{equation}\label{9}
{\bf M. \overrightarrow{x} = \overrightarrow{c}}
\end{equation}
where
\[ {\bf \overrightarrow{c}} = \left( \begin{array}{c} 0 \\ 0 \\ 1 \\1 \end{array} \right) \equiv  \left[ \begin{array}{c}
{\bf 0}  \\ {\bf 1}\end{array} \right]  \]
and
\[ {\bf \overrightarrow{x}} = \left( \begin{array}{c} x_{1} \\ x_{2} \\ x_{3} \\x_{4} \end{array} \right) \equiv  \left[ \begin{array}{c}
{\bf X_{1}}  \\ {\bf X_{2}}\end{array} \right].  \]
Let us denote as $a_{j}\equiv \exp(-i.\zeta_{j}.x) $ and $d_{ij} \equiv \frac{1}{\zeta_{i}-\zeta_{j}^{*}} $ and $d_{ji}^{*} = - d_{ij}$.
The matrix $\bf M$ is defined as:
\[ {\bf M} = \left( \begin{array}{cccc} a_{1} & 0 & \frac{c_{1}^{*}}{a_{1}^{*}}.d_{11} & \frac{c_{2}^{*}}{a_{2}^{*}}.d_{12} \\ 0 & a_{2} & \frac{c_{1}^{*}}{a_{1}^{*}}.d_{21} & \frac{c_{2}^{*}}{a_{2}^{*}}.d_{22} \\ \frac{c_{1}}{a_{1}}.d_{11}& \frac{c_{2}}{a_{2}}.d_{21} & a_{1}^{*}& 0\\
\frac{c_{1}}{a_{1}}.d_{12}& \frac{c_{2}}{a_{2}}.d_{22}   & 0& a_{2}^{*} \end{array} \right) \equiv \]
\[ \equiv  \left[ \begin{array}{cc}
{\bf A} & {\bf B} \\ {\bf -B^{*}} & {\bf A^{*}}  \end{array} \right], \]
where
\[ {\bf X_{1}} = \left( \begin{array}{c} x_{1} \\ x_{2} \end{array} \right), \]
\[ {\bf X_{2}} = \left( \begin{array}{c} x_{3} \\ x_{4} \end{array} \right), \]
\[ {\bf A} = \left( \begin{array}{cc} a_{1} & 0 \\
0 & a_{2} \end{array} \right) \]
and
\[ {\bf B} = \left( \begin{array}{cc} \frac{c_{1}^{*}}{a_{1}^{*}}d_{11} & \frac{c_{2}^{*}}{a_{2}^{*}}d_{12} \\
\frac{c_{1}^{*}}{a_{1}^{*}}d_{21} & \frac{c_{2}^{*}}{a_{2}^{*}}d_{22} \end{array} \right) .\]
The super-matrices and super-vectors are denoted by $\left[ \begin{array}{c}
´ \end{array} \right]$.

The equation (\ref{9}) has the form:
\begin{equation}\label{10}
 \left[ \begin{array}{cc}
{\bf A} & {\bf B} \\ {\bf -B^{*}} & {\bf A^{*}}  \end{array} \right].  
\left[ \begin{array}{c} {\bf X_{1}} \\ {\bf X_{2}} \end{array} \right] =  \left[ \begin{array}{c} {\bf 0} \\ {\bf 1} \end{array} \right], 
\end{equation}
e.i. we obtain in the explicit form:
\[ {\bf A.X_{1} + B.X_{2} = 0}, \]
\[ {\bf -B^{*}.X_{1} + A^{*}.X_{2} = 1}. \]

By solving the system of equations (\ref{10}) by standard method we obtain for the potential $q(x, t)$ (\ref{8}):
 \[ q(x, t) = -2.i [c_{1}^{*}.\exp(i.\omega.x) + c_{2}^{*}.\exp(-i.\omega.x) \] \[ -c_{1}^{*} .\mid c_{2} \mid^{2} .(d_{21}-d_{22})^{2}. \exp(i . \omega .x - 4. \eta.x) - c_{2}^{*} .\mid c_{1} \mid^{2} .(d_{11}-d_{12})^{2}. \exp(-i . \omega .x - 4. \eta.x)]. \]
 \[ [\exp(2.\eta.x) - c_{1}.c_{2}^{*}.d_{12}^{2}.\exp(-2.i.\omega . x - 2.\eta .x)- c_{2}.c_{1}^{*}.d_{21}^{2}.\exp(+2.i.\omega . x - 2.\eta .x) - \]
 \[ \mid c_{2} \mid^{2}. d_{22}^{2}.\exp(-2.\eta.x) - \mid c_{1} \mid^{2}. d_{11}^{2}.\exp(-2.\eta.x) + \mid c_{2} \mid^{2} . \mid c_{1} \mid^{2} . (det(d_{ij}))^{2} . \exp(-6.\eta.x) ]^{-1} , \]
 here $det(d_{ij})$ is the determinant of the matrix with $d_{ij}$ (here $i,j = 1, 2$) elements.
  
  In order $q(x, t)$ to be a real potential, we have to take $q(x, t)^{*}= q(x, t)$, this should we have in mind when choosing $c_{01}$ and $c_{02}$.
  Time dependence will be found when an explicit time dependence of $c_{1}$ and $c_{2}$ will be found, and the relation $\mid \zeta \mid^{2} = \frac{\omega^{2}}{4} + \eta^{2}$ will be taken into account. Then from $q(x, t)$ we will find that the breather solution is:
  \begin{equation}\label{11}
  u(x, t) = 4. \arctan(\frac{2.\eta}{\omega}. \frac{\cos(\omega.x - \frac{\omega}{4.\eta^{2}+ \omega^{2}}.t + \epsilon)}{\cosh(2.\eta.x + \frac{2.\eta}{4.\eta^{2}+ \omega^{2}}.t + \delta)}),
  \end{equation}
  where
  \[ \exp(2.\delta) \equiv \frac{16. \eta^{2}.\mid \zeta \mid^{2}}{\omega^{2}}. \frac{1}{\gamma^{2}} \]
  and
  \[ \exp(2.i. \varepsilon) \equiv \frac{\mid \zeta \mid^{2}}{\zeta^{*2}} .\exp(i. (\beta - \alpha)) \]
  and
  \[ \zeta = i. \eta - \frac{1}{2}. \omega \]
  and
  \[ \mid c_{10} \mid = \mid c_{10} \mid= \gamma\]
  and
  \[ c_{10} = \gamma . \exp(i.\alpha) \]
  and
    \[ c_{20} = \gamma . \exp(i.\beta). \]

    After transition to the $(X, T)$ coordinates, see I. , we obtain:
  \[ u(X, T) = 4. \arctan(\frac{2.\eta}{\omega}. \frac{\cos(X.(\frac{\omega}{2}).(1-\frac{1}{4.\eta^{2}+\omega^{2}})+ T.(\frac{\omega}{2}).(1+\frac{1}{4.\eta^{2}+\omega^{2}})+ \epsilon)}{\cosh(X.(\eta).(1+\frac{1}{4.\eta^{2}+\omega^{2}})+ T.(\eta).(1-\frac{1}{4.\eta^{2}+\omega^{2}})+ \delta)}). \]

  Let us introduce the velocity $U$:
  \[ U \equiv \frac{1-\frac{1}{4.\eta^{2}+\omega^{2}}}{1+\frac{1}{4.\eta^{2}+\omega^{2}}}. \]
  This leads to the relation:
  \[ \frac{1}{4.\eta^{2}+\omega^{2}} = \frac{1-U}{1+U}. \]
  Finally we obtain the solution for the breather:
  \begin{equation}\label{12}
  u(X, T) = 4. \arctan(\frac{2.\eta}{\omega}. \frac{\cos(\frac{\omega}{1+U}(T + X.U)+ \epsilon)}{\cosh(\frac{X+U.T}{\sqrt{1-U^{2}}}.\sqrt{1 - \frac{\omega^{2}}{1+U}}+ \delta)}).
    \end{equation}
    Note that $2.\eta = \sqrt{\frac{1+U}{1-U}- \omega^{2}}.$
    In the eigencoordinate system we have $U=0$ and the equation (\ref{12}) gives reference (5) from I. . The zero value of $U=0$ corresponds to $4.\eta^{2}+\omega^{2}=1,$ e.i. $\mid \zeta \mid^{2} = \frac{1}{4}.$
\section{Discussion.}
It the paper was discussed the Inverse Scattering Method used as a methodological Part II to the Part I. from the paper I.  to solve the Sine-Gordon Equation and to find the breather solution. While the aim of the paper I. was to introduce the Inverse Scattering Method for later studies of some problems in nonlinear dynamics, as a methodological example we described how to solve the Sine-Gordon Equation using the Inverse Scattering Method to obtain a soliton (kink). In this paper we discussed another methodological example: the solution of the Sine-Gordon Equation using the Inverse Scattering Method to obtain description of a breather. We will discuss later breather solutions in some physical systems.
This is the reason why we do not discuss mathematical properties of the kink and of the breather, and (for an physical system) physical properties of the kink and of the breather. Thus we have in the Part I. and in this Part II. seeds for solving problems using the Inverse Scattering Method in general, and  in the case of the Sine-Gordon Equation  we have found solutions describing the soliton and the breather.

\end{document}